# SONIA: an immersive customizable virtual reality system for the education and exploration of brain networks

Owen Hellum[1], Christopher Steele[2,3,4], *Yiming Xiao[1,3]

[1]Department of Computer Science and Software Engineering, Concordia University, Montreal, Quebec, Canada

[2]Department of Psychology, Concordia University, Montreal, Quebec, Canada

[3]PERFORM Centre, Concordia University, Montreal, Quebec, Canada

[4]Department of Neurology, Max Planck Institute for Human Cognitive and Brain Sciences, Leipzig, Germany

**Running title**: *SONIA: VR for brain network education*

**Corresponding Author**

**Yiming Xiao**
Department of Computer Science and Software Engineering
Concordia University
1455 Boulevard de Maisonneuve O.,
Montreal, Quebec
Canada H3G 1M8

E-mail: yiming.xiao@concordia.ca

# Abstract

While mastery of neuroanatomy is important for the investigation of the brain, there is an increasing interest in exploring the neural pathways to better understand the roles of neural circuitry in brain functions. To tackle the limitations of traditional 2D-display-based neuronavigation software in intuitively visualizing complex 3D anatomies, several virtual reality (VR) and augmented reality (AR) solutions have been proposed to facilitate neuroanatomical education. However, with the increasing knowledge on brain connectivity and the functioning of the sub-systems, there is still a lack of similar software solutions for the education and exploration of these topics, which demand more elaborate visualization and interaction strategies. To address this gap, we designed the immerSive custOmizable Neuro learnIng plAform (SONIA), a novel user-friendly VR software system with a multi-scale interaction paradigm that allowed flexible customization of learning materials. With both quantitative and qualitative evaluations through user studies, the proposed system was shown to have high usability, attractive visual design, and good educational value. As the first immersive system that integrated customizable design and detailed narratives of the brain sub-systems for the education of neuroanatomy and brain connectivity, SONIA showcased new potential directions and provided valuable insights regarding medical learning and exploration in VR.

# Introduction

The human brain is a highly complex organ that consists of small anatomical structures that are tightly packed and interconnected through different pathways. To aid spatial understanding and exploration of the brain's 3D anatomy, volumetric data is often sliced into 2D representation due to the limitations of traditional media (e.g., paper and 2D screens). However, this often fails to effectively reflect the complex geometry and spatial arrangement of the anatomical structures (Ekstrand et al., 2018; Xiao et al., 2018). With the advancement of modern bioimaging techniques, the exploration of functional and structural brain connectivity is gaining increasing interest. Intuitive demonstration of brain connectivity along the associated neuroanatomy and the insights gained through various studies will be instrumental to the education and further exploration of neuroscience (Petersen et al., 2019). So far, a number of augmented reality (AR) and virtual reality (VR) solutions (Fiani et al., 2020; Hellum et al., 2022b) have been proposed to provide more intuitive visualization and understanding of neuroanatomy for educational and surgical planning purposes, with positive responses from user studies. These solutions have employed a range of display devices, including mobile devices (e.g., tablet and smartphone), VR headset, and Hololens. In comparison to the primary focus on the anatomy, only a few AR/VR systems (Karmonik et al., 2018; Keiriz et al., 2018; Petersen, et al., 2019; Schloss et al., 2021; Hellum et al., 2022a) have been proposed to visualize and demonstrate the neural pathways and brain networks. Keiriz, et al. (2018) proposed NeuroCave, a web-based immersive platform for exploring connectomic data. Later, workflows that leverage existing software solutions to visualize brain tractograms and functional connectivities have been demonstrated (Karmonik, et al., 2018; Petersen, et al., 2019). More recently, Schloss, et al. (2021) built a VR application to visualize the information pathways of visual and auditory systems for educational purposes. While existing solutions tackle the challenges in spatial understanding of the 3D anatomy through visualization, very few experimented with new user interaction paradigms, which can potentially enhance the usability and learning experience (Hellum, et al., 2022a). In addition, among the limited efforts (Karmonik, et al., 2018; Keiriz, et al., 2018; Petersen, et al., 2019; Schloss, et al., 2021; Hellum, et al., 2022a) in visualizing brain networks, no reports attempted to incorporate descriptive insights along the pathway exploration or learning module design.

To meet the emerging need for the education, demonstration, and investigation of brain connectivity and to promote related neuroscientific insights, we proposed the immerSive custOmizable Neuro learnIng plAform (SONIA), which provides interactive visualization and learning modules for both neuroanatomy and the associated structural and functional networks. The new VR system has several novel features. First, inspired from VR-based geological data navigation (Piumsomboon et al., 2018; Huang and Chen, 2019), we experimented with a multi-scale interaction paradigm that places the user at the centre of a large, expanded brain while also manipulating a small brain model to facilitate spatial understanding of brain anatomy. Second, we designed a progression-based strategy with completion metrics and multimedia interactions, including visual guidance and audio voice-over to provide a stimulating and enriching user experience. Finally, the system's customizable design to incorporate detailed narratives of brain sub-systems opens the door for future projects, allowing many different types of content to be visualized and explored with the proposed software framework. To demonstrate the proposed system, we created an interactive visualization of the research work of Xie et al. (2021) on the functional system and brain network of anxiety. We conducted quantitative and qualitative user assessments that indicated that the system exhibits excellent usability, visual design, and educational value. Thus, in conjunction with conventional learning materials composed of 2D graphic representations, our proposed novel, customizable, and intuitive VR system has significant promise and value for the education and exploration of neuroanatomy and neural pathways. The code of the SONIA system is made publicly available at https://github.com/HealthX-Lab/SONIA.

## Materials and Methods

### Virtual brain model

To demonstrate the proposed VR system, we used the anxiety-relevant functional brain network summarized in a recent review by Xie, et al. (2021). A summary of the network, which involves six key anatomical structures (amygdala, hippocampus, striatum, medial prefrontal cortex (mPFC), hypothalamus, and the bed nucleus of the stria terminalis (BNST)) is illustrated in Fig 1 of the paper by Xie, et al. (2021). Briefly, Xie *et al*. summarized five subsystems that regulate anxiety, including cognitive control, fear conditioning, uncertainty anticipation, motivation processing, and stress regulation; with each subsystem made up of pathways between two to three anatomical structures. For the system, we constructed the virtual brain model based on the AAL116 brain atlas (Tzourio-Mazoyer et al., 2002), which is widely used in neuroimaging research. Five of the six key structures involved in the anxiety-related functional systems were extracted from the atlas. As the AAL116 atlas does not contain the BNST, it was segmented manually according to Theiss et al. (2017) on the MNI152 brain template (Fonov et al., 2011) in the same space as the AAL116 atlas. All atlas structures were converted to .fbx mesh models from the discrete labels in the NIFTI images for use in the 3D VR environment. For the virtual brain models, we only highlight these six structures while keeping the rest as semi-transparent to provide the additional visual references to further enhance the spatial understanding of the anatomy and richness in the final rendering. Finally, both opaque and semi-transparent lines were added between both the six key structures and the rest of structures from the AAL116 atlas, respectively, to indicate functional and anatomical connectivities between them.

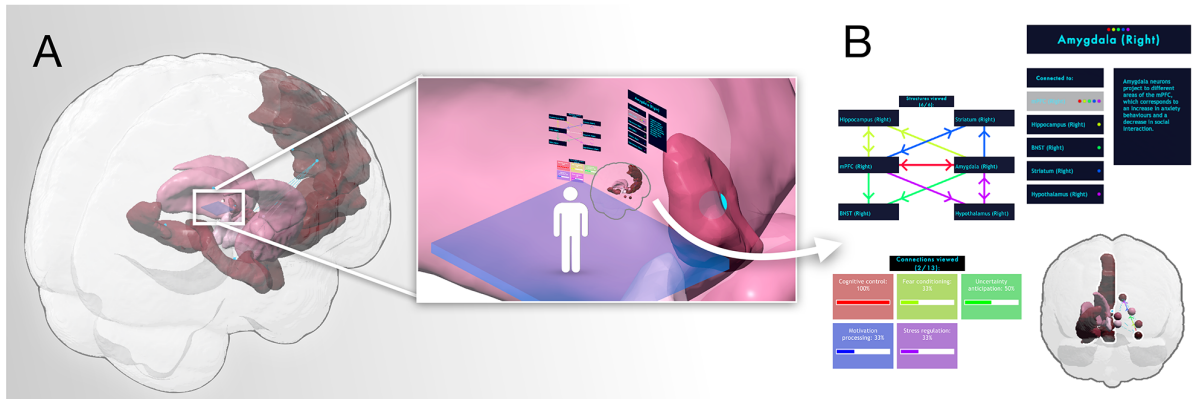

**Figure 1**. Overview of the virtual reality environment set-up. **A**. The spatial relationship between the "large brain" environment and the "mission control" platform; **B**. The layout of the user-interface, which consists of three information panels and a small brain model that allows the user to interact with the "large brain" environment and complete the learning modules.

**Virtual reality environment construction**

For the virtual environment, we explored a multi-scale visualization paradigm with two brain models of different sizes to facilitate interaction, visualization, and spatial understanding of the neuroanatomy. Multi-scale VR interaction was recently suggested for the exploration and navigation of geographic data to facilitate the understanding of spatial arrangement (Piumsomboon, et al., 2018; Huang and Chen, 2019). For anatomical navigation, we expect that this approach will also benefit the spatial understanding of the neuroanatomy, as well as provide an enriching, fun, and immersive experience for the user.

In the VR environment, the user is positioned on a "mission control" platform suspended at the center of a magnified brain model at the scale of a large house, which is out of reach for the user, but still allows clear recognition of the spatial arrangement of different anatomical structures (hereafter referred to as the large brain). At the same time, a smaller forward-facing brain model that mirrors the large brain is placed in front of the user to allow interaction with the learning modules and the large brain model (the small brain). Along with the small brain, three floating information panels are also presented to the user to display the schematic diagram of anxiety-related functional subsystems, descriptions for all brain connectivities, and the percentage of completion for the learning content for each functional sub-system. The schematic of the designed VR environment is demonstrated in **Fig. 1** and the details of each visual element for the "mission control" platform are illustrated in **Fig. 2**. Different from the magnified brain, the small brain displays the geometries of the six key anatomical structures with different shades of red on the left hemisphere while the right hemisphere depicts a graph representation such that each region is a color-coded sphere (located in the regional centroid) and their connectivities are denoted by connecting lines. The node in this graph representation offers clear visualization of the connectivity relationships between the anatomies and makes it easier to select each anatomical structure, allowing the small brain model to serve as the main media to interact with the rest of the visual elements in the virtual environment. Corresponding to the right hemisphere of the small brain model, lines that connect the centroids of the key structures are also shown on the same side of the magnified brain.

Our VR system was created using the Unity game engine (version 2021.3.2f1) with the SteamVR plugin. We employed the HTC VIVE Pro Eye VR HMD headset and a Razer Blade 15 laptop (Intel Core i7 CPU, NVIDIA GeForce RTX 2070 GPU, and 16 GB RAM) to run the system. No lagging or frame freezing was observed for our system, and it ran consistently at an average of 45-50 frames per second. Only one VR controller is required to perform target selection and confirmation for the VR system.

**Overview of the system workflow**

Before understanding the subsystems and brain networks that regulate a neural process, it is important to first elucidate the spatial arrangement of each neuroanatomy that is involved. Therefore, we designed the workflow for the user in the SONIA system in two general phases (anatomical learning and connectivity learning), both utilizing a single VR controller in the dominant hand for pointing and selection. In the first phase, the user is guided to learn about the key brain structures involved in the target neural network. Upon completion, the user is guided to the second phase to explore the connections between the structures and the roles of different subsystems in a neural process, until all subsystems have been visited. At each step of the workflow, we have designed appropriate user interaction strategies that fully utilize the visual elements in the environment to provide a stimulating experience. In both stages of the system, the user does not need to select the structures and connections in any predefined order, thus giving them the opportunity to select items and knowledge points that most interest them, or where perhaps closely related to the structures that they had just visited. By granting participants this freedom, the users are given a chance to exercise limited agency in their own educational experiences and learn at their own pace. We will further elaborate the user interaction strategies and system workflows for the two phases in the following sections.

**Anatomical learning phase**

During anatomical learning, the user is tasked to navigate all key brain structures to learn their spatial arrangements and roles in the neurological system. To accomplish this, a short virtual stick extends from the controller with a small sphere at the tip, which is used as the default pointing and selection tool. The workflow of this phase is illustrated in **Fig. 3**. When the user touches the target object with the virtual stick, the hit object becomes highlighted with a white halo, which, when confirmed by pressing the controller's trigger button, will remain to highlight the structure. This user interaction strategy is only applicable on the right hemisphere of the small brain model, where the key anatomical structures are represented as interconnected nodes for selection to reduce visual clutter. Once a node selection is confirmed, the corresponding anatomical structure in its full geometric representation in the left hemisphere of the small brain becomes highlighted with a white halo. Syncing with the interaction upon the small brain, the same white halo that indicates selection and hover-over is also shown in the corresponding structures in the right side of the large brain. This signals the user of the link between the two brain models of different scales.

Two information panels are employed as the key UI elements in anatomical learning. First, the learning material panel is positioned above the small brain to display the name and the key knowledge points for the selected brain anatomy. Second, the connectivity diagram (**Fig. 2C**) demonstrating the relationship between the anatomical structures is placed to the left of the small brain. Although the panel is empty at the beginning, once a structure is visited, the corresponding item will become visible in the diagram until all structures have been selected at least once. The connections between structures will also be revealed as the item list populates.

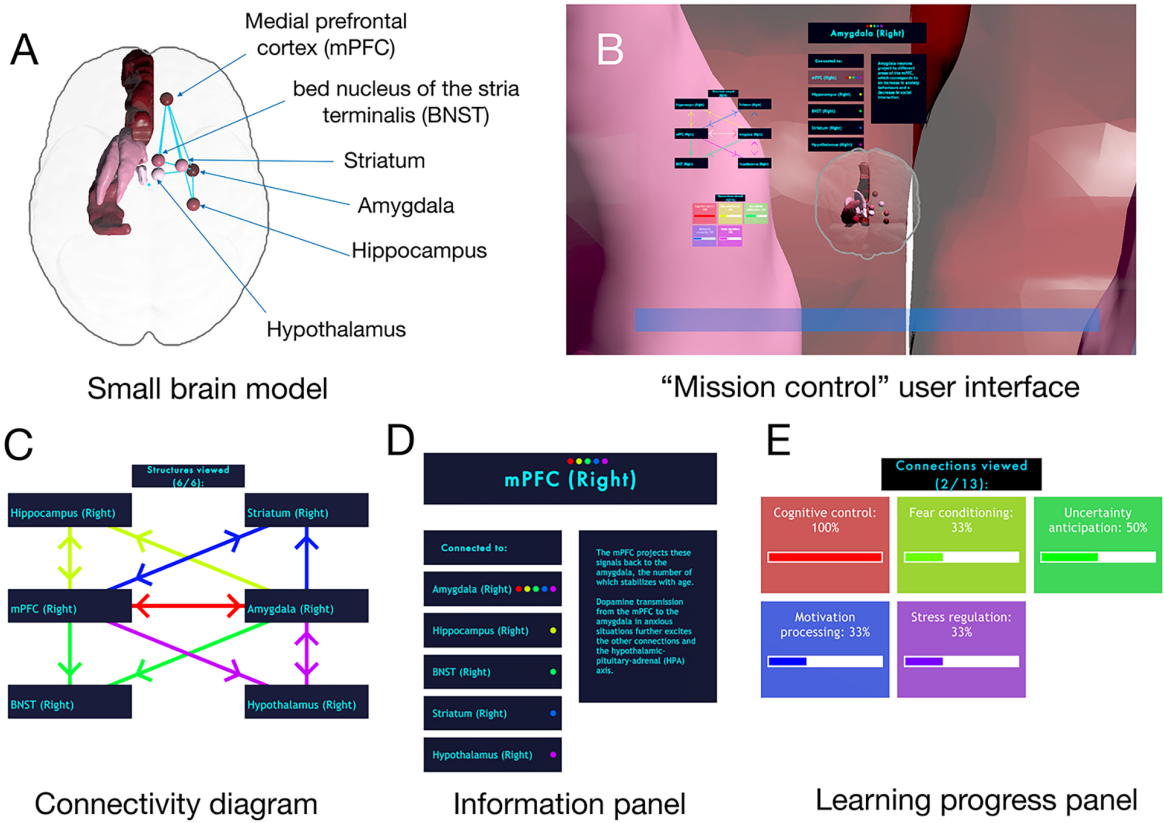

**Figure 2**. Detailed demonstration for the user-interface of SONIA. **A**. Composition of the small brain model that allows interaction and visualization of brain anatomy and connectivity; **B**. Inside view of the "mission control" platform; **C.** Schematic diagram of anxiety-related functional systems and brain structures; **D**. Information panel for displaying learning materials regarding the key brain regions shown in **C**; **E**. Information panel that notifies learning material progress. Note that across **C**, **D**, and **E**, the same color-coding strategies are used consistently to code for different processes in the response to anxiety.

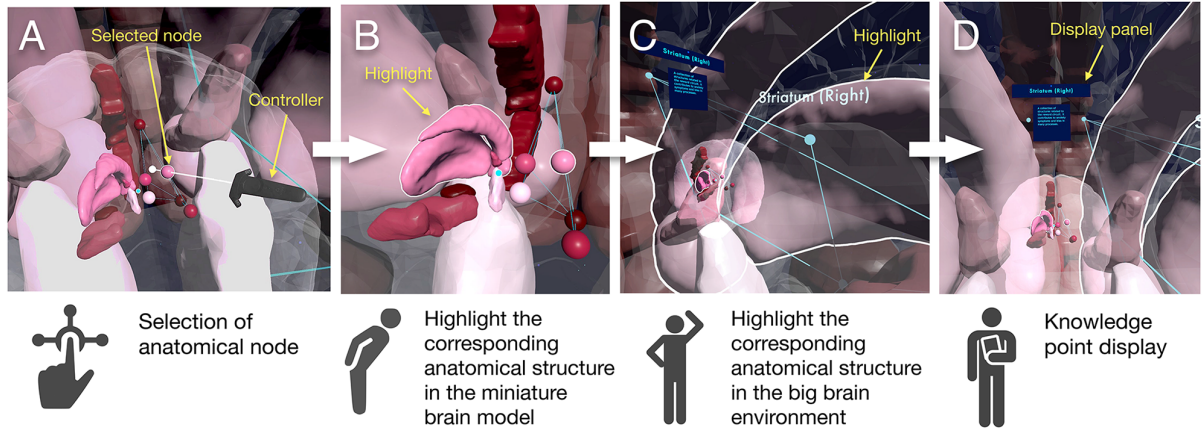

**Figure 3.** Workflow for anatomical learning phase. **A.** the user points the controller line into a structure and presses the trigger to select it, **B.** the structure becomes highlighted in the small brain, **C.** the structure becomes highlighted in the large brain (background), **D.** the display panel showing the name and description of the selected structure.

**Connectivity learning phase**

Upon completing the anatomical learning module, the user will proceed to the connectivity learning phase, where all three information panels illustrated in **Fig. 2C-2E** are employed together with the small and large brain models to fulfill an interactive learning experience. Note that among the three panels, only the one that displays the learning materials has selectable menus for direct user interaction, and its setup is different from that of the anatomical learning. While we maintain the user interaction strategy in Section 2.4 for picking anatomical structure using the small brain, a 'laser pointer' now extends from the controller which is used to select menu items in the learning material panel, as it is not within the arm's reach.

To start investigating a connection, the user needs to first pick a brain structure from the small brain model. Then, the name of this structure and those in the network that it passes information towards will be listed in the learning material panel. For each menu item that represents a unidirectional connection, small dots with the color-coding that signifies the membership of a subsystem are marked on it as well. Then, a further selection of an item in the list will trigger the display of the key knowledge points regarding the description of the connectivity between the two structures within the brain network under study. At the same time, this connection under investigation and its directionality will be annotated in both the large and small brain models using colour coding strategies corresponding to the subsystem(s) that it belonged to, as well as in the connectivity diagram (**Fig. 2C**) using a white color. As the user gradually explores all connections in the connectivity diagram, the progress panel will track the completion of the learning materials for each subsystem with bar graphs showing the percentages of the connections that have been viewed. To better demonstrate the workflow, an example of exploring the connection from the amygdala to the mPFC using the SONIA system is shown in **Fig. 4**.

As mentioned previously, colour coding is used extensively throughout the experience, such as on all the information panels, both to denote a belonging to a particular structural subsystem and to show which structures and/or connections are selected. As a part of the customizable design, visually distinct colours are automatically generated by SONIA for each of the system's subsystems. The use of colour coding facilitates the user to establish immediate association between the connections and their subsystems. Note that the white color is reserved for our software system to indicate that the structure and/or connections have been selected.

**Customizable system design**

To enable flexibility and adaptability for new learning materials, our proposed SONIA system was designed in such a way as to allow alternative datasets that define the anatomical models and the functional relationship between them to be loaded. Specifically, the following data are necessary for the system to function: a collection of 3D model files (e.g., .fbx, .obj, etc.) for the anatomical structures, a .csv spreadsheet containing the names and descriptions of the structures, and a .csv spreadsheet with the connectivity matrix between structures. Additional files are optional but can further enhance the learning experience. They include .csv files that list the subsystem names and descriptions, membership of structures and connections to the subsystems, as well as extra 3D model files for peripheral anatomical structures and their connectivity matrix to help enrich the visual content if needed. Besides these customizable data for alternative learning modules, the users are also welcome to tweak the visualization styles (e.g. colors and mesh textures) in the Unity editor. As the Unity scenes, user interaction strategies, and UI displays are programmed, they will remain unchanged in customization. Furthermore, to achieve the optimal visualization of the information panel for displaying connectivity diagrams, the user will be encouraged to design the layout that best suits their target applications and population. By placing these required files in a specific folder and updating the editor script variables to point to the correct locations and change any additional settings, different learning contents can be generated for either subject-specific brain models or existing brain atlases (e.g., AAL116). With even a simple set of meshes, a connectivity map, and structure descriptions, an interactable experience can be produced. By leveraging the AAL116 atlas, our demonstrated case study took full advantage of such a setup. As no frame rate loss or system errors were observed with full rendering of all brain parcellations of the atlas with different levels of transparency, we believe that the system is highly scalable for complex neuroanatomical models.

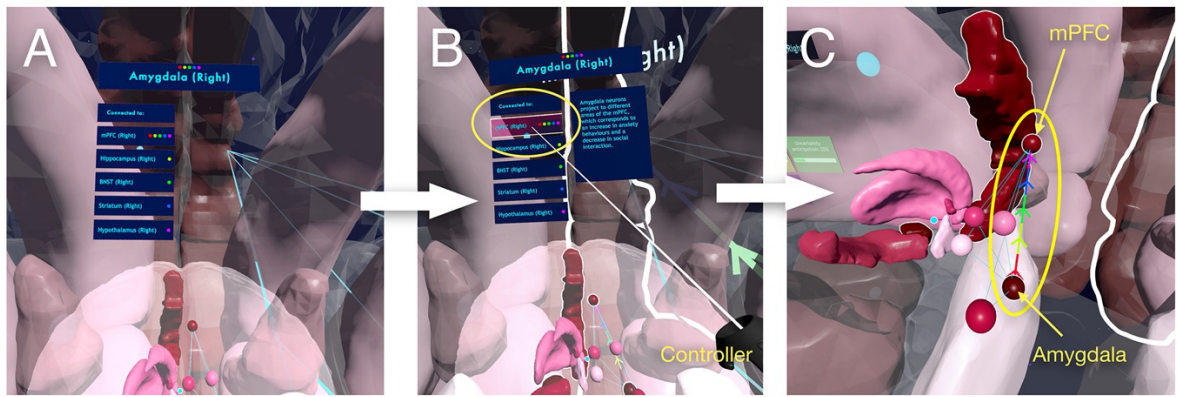

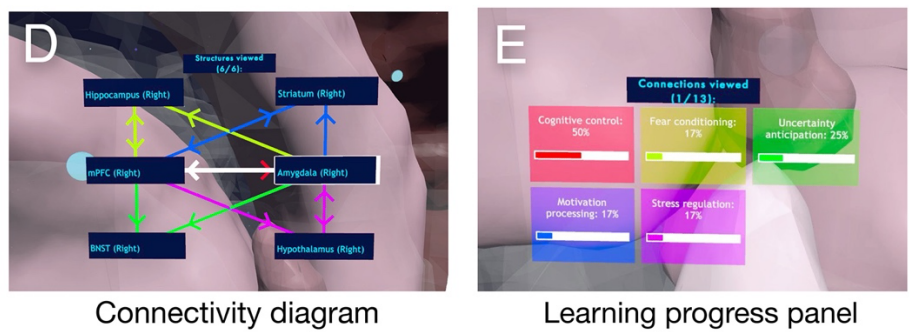

**Figure 4.** Workflow of the connectivity learning phase using the connection from the amygdala to the mPFC as an example (yellow circles indicate important events). **A**. Information panel showing all available brain structures that the amygdala is connected to when it is selected by the user. **B**. When the mPFC is selected, the description of the connection in anxiety processing is shown. **C.** the connection becomes highlighted in the small (and the large) brain. **D**. connectivity diagram, with the currently selected connection highlighted in white. **E.** subsystem completion diagram with percentages of completion for each subsystem for anxiety processing. Please note that all the arrows indicating the directions of the connectivity are color-coded by the corresponding subsystems.

**User study design and system validation**

The usability of the SONIA system was assessed with both quantitative and qualitative evaluations in user studies. Upon informed consent, we recruited 11 subjects (age=31.1±6.0, 4 female, 7 male) to participate in the study. All participants were either somewhat or very familiar with neuroanatomy and/or the concept of brain connectivity, and represent the main target users of the system. Among them, only one did not have VR experience before the study. The study was approved by the Ethics Research Board of Concordia University. All methods were performed in accordance with the relevant guidelines and regulations.

Subjects spent 20~30 minutes following the directions of a tutorial while they interacted with the environment. The tutorial consisted of a text-to-speech (TTS) voice-over and accompanying texts that described the interactions and responses of the system. This tutorial guided participants through each interaction scheme and visual change, and explained all the subsystems within the loaded data. Subjects with glasses were allowed to wear them while participating, as the HTC VIVE Pro Eye headset is compatible with them. No participants experienced motion sickness.

Upon completing the VR experience, we asked each participant to complete a three-part questionnaire, consisting of both quantitative and qualitative assessments. The first part of the questionnaire contained the System Usability Scale (SUS) evaluation (Brooke, 1995), which is widely used to validate the usability of software systems. It is comprised of 10 questions on the scale of 1-5, evaluating complexity, user-friendliness, and confidence when using a software system. Among the 10 questions, each odd-numbered question is scored as x-1, and each even-numbered question is scored as 5-x, where x is the question's resulting value. The scores for each participant are then summed, and then multiplied by 2.5 - resulting in a maximum SUS score of 100. A software system that receives a SUS score above 68 indicates good usability. The second part of the questionnaire contained an additional feedback form pertaining to the visual design, interaction design, and learning experience of the system (rated along scales from 1-5 for each sub-question, with higher values indicating more positive responses). They complement the SUS results to further identify the effectiveness of our system's more specific design decisions. First, for visual design, three questions were asked to establish the perceived complexity of the visual elements, pleasantness of graphic styles, and usefulness of colour coding schemes for sub-system representation. They assess whether we have appropriately balanced the accuracy of data representation in the UI and the artistic colours and placements of visual elements. The averaged score of these questions was also obtained to obtain an overall assessment of the visual design. Second, an interaction design question asked participants how well the interface communicated the procession of navigating through the anatomical structures and neural connectivity. Thirdly, the learning experience had two questions to determine whether the multi-scale navigation strategy with the small and large brains enhances spatial understanding of the anatomy and the overall learning yield of the system. For the total SUS score, a one-sample t-test was used to assess whether the results are significantly different from 68, and for each sub-score in Part 1 and 2, we compared the results to a neutral response (score=3), also with one-sample t-tests. Here, a p-value<0.05 indicates a statistically significant difference. Finally, in Part 3 of the questionnaire, we provided open-ended questions to allow the participants to provide additional suggestions on how to improve the system further and justify some of the ratings given to the previous sections as they see fit. The responses were reviewed carefully to help understand the quantitative assessments and potential directions for future improvements.

## Results

### Quantitative evaluation

The full SUS score from the user study was 79.8±11.6, significantly greater than the minimum 68 for a usable system ($p=0.007$). Every sub-score, except system complexity, indicated a positive user-interaction ($p<0.05$) with the SONIA system for the overall ease of use, intuitiveness, and consistency. Interestingly, the opinions of the participants were divided in terms of the complexity of the system, resulting in a rating of 2.3±1.2 (lower score indicates lower perceived system complexity), though this value was not statistically different from a neutral response (p=0.07).

In Part 2 of the questionnaire, the feedback for visual, interaction, and learning experience was generally positive. While the overall visual score was 3.9±0.5 (p=0.0001), for three associated sub-scores, the results for the complexity of the visual elements, pleasantness of graphic styles, and usefulness of colour-coding schemes for sub-system representation were 3.5±1.1 (p=0.14), 4.2±0.8 (p=0.0004), and 4.0±0.9 (p=0.004) respectively. With a score of 3.6±0.9, the interaction design was rated as effective for navigating the anatomies and connections (p=0.046). Finally, though participants were mostly neutral (3.1±1.1, p=0.80) regarding the multi-scale strategy for enhancing anatomical understanding, they felt strongly that they had learnt a lot (3.9±0.7, p=0.002) while using the system.

**Qualitative evaluation**

Besides quantitative responses, the feedback form also contained a qualitative section to allow participants to remark freely upon general impressions, opinions, and improvements regarding the system. Participants commented about both positive and negative aspects of their experience. In particular, users found the utility and function of the system to be helpful and novel, standing out as a good way to represent the data. In terms of the user interface, participants often found it to be containing too much visual information. Due to the large volume of text and UI placements, they felt mildly overwhelmed, and found it difficult to absorb some of the informative material.

## Discussion

Using the functional network of anxiety regulation (Xie, et al., 2021) as a case study, the proposed SONIA VR system is the first to integrate descriptive insights along neural pathway exploration and learning module design. Besides visualization of the 3D anatomy, we explored novel interaction and user-interface designs intended to benefit the usability and user experience. As the knowledge of neuroanatomy is a prerequisite to understanding neural pathways, we designed the workflow of the system to encompass the phases of anatomical and connectivity learning. In each phase, following the popular practice of player agency (a practice in game design to leverage control of the environment towards the player), the user is free to select the anatomies and the associated connections at their will and pace to trigger changes in the UI elements in the virtual environment. Together with the information panel that displays the progress of completion, these components are designed to enhance the motivation and ease of using the system. Both strategies have shown positive impacts in the design of games and educational content (Plass et al., 2015; Taub et al., 2020). The system focused on this built-in reward system (the visual demonstration of progress), rather than point- or trophy-based rewards, as these less tangible markers of completion have been shown to be less impactful on feelings of reward and on learning amount (McKernan et al., 2015). The positive feedback from our user evaluation regarding the willingness to use the system frequently partially signifies the benefit of these user-interaction designs.

Besides leveraging the superiority in 3D visualization of virtual reality (Ekstrand, et al., 2018), the ease of use of our system is key to facilitating the understanding of complex neural pathways in the brain. To achieve this, we implemented a number of visualization and interaction strategies. First, the small brain model is used as the central device to interact with the rest of the environment and UI elements; Second, both node-based and full anatomical representations are used to reduce clutter, facilitate spatial understanding, and simplify object selection; finally, systematic colour-coding is employed to signify the association to the subsystems of the brain network. Their positive impacts are confirmed by the SUS assessments, particularly in the sub-scores for the intuitiveness and ease of use, as well as the visualization and interaction experience evaluations. In terms of the complexity of both the software design and visual elements, the participants indicated slightly favorable (but not statistically significant) opinions, respectively. The discrepancy may be subject to the participants' varied levels of experience with the VR system and neuroscience. This leaves more room to improve the system further in our future strategies. Potential solutions could involve further simplification of the UI and options to expand descriptions rather than have them be constantly present - these changes would serve to reduce the amount of textual information presented to the user at any given time. Another interesting fact is that the participants reported that they were neutral on the beneficial role of multi-scale representation of the brain for anatomical understanding. This may be due to the choice of scales and the limited freedom of active movement in the large brain model, which is in contrast to previous works in geographic data exploration, where multiscale approaches have been shown to be beneficial (Piumsomboon, et al., 2018; Huang and Chen, 2019). However, as the representation forms the overall visual style of the system, creating a visually appealing and enriching environment, the overall visual styles were highly appreciated by the participants. In addition, the user study confirmed a highly positive learning experience from the participants.

In the freeform feedback, most participants (9/11) listed a small number of remarks on both the usefulness and difficulties that they encountered with the system, as well as suggestions for improvements. Among them, the issue of complexity of the user interface and the large volume of knowledge points was mentioned by several participants (5/11), but none reported being unable to learn or being too overwhelmed. Although the multi-scale strategy for spatial learning was rated neutrally, a subset of participants (4/11) reported liking it and praised the system as a learning tool. The more detailed responses show that although improvements are still needed to better present the complex information, which is indeed challenging as suggested by previous works (Keiriz, et al., 2018; Ille et al., 2021; Schloss, et al., 2021), the overall system was generally well received.

Due to the time constraint, we showcased here the proposed system with only a single example of brain pathways as a proof of concept. However, based on the results of the user studies, it has demonstrated good usability, positive user experience, and educational value. With the customizable design, which supports easy adaptation of alternative learning content, we will continue to evaluate the performance and impact of the software platform with new materials and network models. As brain connectivity is a more advanced topic than neuroanatomy and the main focus of SONIA, our target user group (and the recruited participants in the user study) are those who are at least somewhat familiar with neuroanatomy. Along with other suggestions of improvement from the user study, further strategies for the anatomical learning phase will be developed and tested for lay users in the future.

## Conclusion

We have built a novel virtual-reality system, SONIA, with a customizable design to help create an immersive learning experience for understanding and demonstration of functional brain systems and networks. Unlike the prior arts that primarily focus on simple anatomical visualization, our proposed system integrates a more immersive, user-friendly, and enriching environment with detailed narratives of the brain sub-systems and effective user-interaction strategies, which is validated through user studies. Through this prototype as the first system of its kind, we demonstrate new potential directions regarding medical learning and exploration in VR.

## Acknowledgements

The authors thank the insightful discussion on brain networks and VR user interaction design with Dr. Najmeh Khalili-Mahani.

## Notes on Contributors

Owen Hellum, B.Sc., is a 4th-year Bachelor of Computer Science student in the Computation Arts program and a member of the Technoculture, Arts and Games Institute of Concordia University, Montreal, Canada. His research work involves the development of virtual reality planning and education systems for neurosurgery.

Christopher Steele, Ph.D., is an Assistant Professor at the Department of Psychology of Concordia University, Montreal, Canada and an adjunct member of the Department of Neurology, Max Planck Institute for Human Cognitive and Brain Sciences, Leipzig, Germany. His research interests focus on neural plasticity, motor learning, and diffusion MRI.

Yiming Xiao, Ph.D., is an Assistant Professor at the Department of Computer Science and Software Engineering of Concordia University, Montreal, Canada. His research combines novel techniques in medical imaging principles, computer vision, and machine learning to improve the efficiency and accuracy of image-based diagnosis and medical procedures.